 \documentclass{aa}
\usepackage[latin1]{inputenc}
\usepackage{natbib}
\usepackage{theorem,algorithm,algorithmic}
\usepackage{amssymb,amsmath}
\usepackage{setspace,ulem}

\usepackage{psfig,epsfig,amssymb,alltt}
\usepackage{latexsym}
\usepackage{graphicx}
\graphicspath{{PS/}}
\usepackage{psfig,amssymb,alltt}

\psfigurepath{./PS:.}
\usepackage[english]{babel}
\usepackage{indentfirst}
\psfigurepath{./PS}
\vspace{1cm}

\newcommand{\JF}[1]{{#1}}
\newcommand{\SD}[1]{{#1}}

\newcommand{\var}[1]{{\mathrm{Var}}\left[#1\right]}

\newcommand{\parenth}[1]{\left({#1}\right)}
\newcommand{\crochets}[1]{\left[{#1}\right]}

\newcommand{\be}{\begin{eqnarray}}
\newcommand{\ee}{\end{eqnarray}}

\newcommand{\bitem}{\begin{itemize}}
\newcommand{\eitem}{\end{itemize}}

\newcommand{\bE}{\mathbb{E}}

\newcommand{\bZ}{\mathbb{Z}}

\newcommand{\fX}{\mathbf{X}}

\psfigurepath{./PS}

\title{Source detection using a 3D sparse representation: application to the Fermi gamma-ray space telescope}

\author{ J.-L. Starck\inst{1}, J.M. Fadili\inst{2}, S. Digel\inst{3},  B. Zhang\inst{4} and J. Chiang\inst{3} }

 \institute{ 
\inst{1} CEA, IRFU, SEDI-SAP,  Laboratoire Astrophysique des Interactions Multi-\'echelles (UMR 7158) , 
CEA/DSM-CNRS-Universite Paris Diderot, Centre de Saclay,  F-91191 Gif-Sur-Yvette, France. \\
\inst{2} GREYC CNRS UMR 6072, Image Processing Group, ENSICAEN 14050, Caen Cedex, France. \\ 
\inst{3}  Stanford Linear Accelerator Center \& Kavli Institute for Particle Astrophysics and Cosmology,  Stanford, CA 94075, USA.  \\
\inst{4}  Quantitative Image Analysis Unit URA CNRS 2582, Institut Pasteur, 25-28, Rue du Docteur Roux, 75724 Paris Cedex 15. \\
}

\offprints{jstarck@cea.fr}

\date{\today}
\authorrunning{Starck et al}
\titlerunning{Source Detection and Fermi Telescope}

\begin{document} 

\abstract{
The multiscale variance stabilization Transform (\JF{MSVST})  has recently been proposed for Poisson data denoising \JF{\citep{starck:zhang07}}. This procedure, which is nonparametric, is based on thresholding wavelet coefficients. The restoration algorithm applied after thresholding provides good conservation of source flux. 
 We present  in this paper an  extension of the \JF{MSVST} to 3D data\JF{{\textemdash}in fact 2D-1D data{\textemdash}} when the third dimension is not a spatial dimension, but the wavelength, the energy, or the time.
We show that the \JF{MSVST} can be used for   detecting and characterizing  astrophysical sources of high-energy gamma rays, using realistic simulated observations with the Large Area Telescope (LAT).  The LAT was launched in June 2008 on the Fermi Gamma-ray Space Telescope mission.  Source detection in the LAT data is complicated by the low fluxes of point sources relative to the diffuse celestial foreground, the limited angular resolution, and the tremendous variation in that resolution with energy (from tens of degrees at $\sim$30 MeV to $\sim$0.1$^\circ$ at 10 GeV).  The high-energy gamma-ray sky is also quite dynamic, with a large population of sources such active galaxies with accretion-powered black holes producing high-energy jets, episodically flaring.  The fluxes of these sources can change by an order of magnitude or more on time scales of hours.  Perhaps the majority of blazars will have average fluxes that are too low to be detected but could be found during the hours or days that they are flaring.

The \JF{MSVST} algorithm is very fast relative to traditional likelihood model fitting, and permits efficient detection across the time dimension and immediate estimation of spectral properties.  Astrophysical sources of gamma rays, especially active galaxies, are typically quite variable, and our current work may lead to a reliable method to quickly characterize 
the flaring properties of newly-detected sources.
}

\maketitle



\section{Introduction}

The high-energy gamma-ray sky will be studied with unprecedented sensitivity by the Large Area Telescope (LAT),  which was launched by NASA on the {\textit{Fermi}} mission in June 2008.  
The catalog of gamma-ray sources from the previous mission in this energy range, EGRET on the Compton Gamma-Ray Observatory, has approximately 270 sources~\citep{egret:hartman99}.  For the LAT, several thousand gamma-ray sources are expected to be detected, with much more accurately determined locations, spectra, and light curves.

We would like to reliably detect as many celestial sources of gamma rays as possible.  The question is not simply one of building up adequate statistics by increasing exposure times.  The majority of the sources that the LAT will detect are likely to be gamma-ray blazars (distant galaxies whose gamma-ray emission is powered by accretion onto supermassive black holes), which are intrinsically variable.  They flare episodically in gamma rays.  The time scales of flares, which can increase the flux by a factor of 10 or more, can be minutes to weeks. The duty cycle of flaring in gamma rays is not well determined yet, but individual blazars can go months or years between flares and in general we will not know in advance where on the sky the sources will be found.

The fluxes of celestial gamma rays are low, especially relative to the $\sim$1 m$^2$ effective area of the LAT (by far the largest effective collecting area ever in the GeV range). An additional complicating factor is that diffuse emission from the Milky Way itself (which originates in cosmic-ray interactions with interstellar gas and radiation) makes a relatively intense, structured foreground emission.  The few very brightest gamma-ray sources will provide approximately 1 detected gamma ray per minute when they are in the field of view of the LAT.  The diffuse emission of the Milky Way will provide about 2 gamma rays per second, distributed over the $\sim$2 sr field of view.

For previous high-energy gamma-ray missions, the standard method of source detection has been model fitting --- maximizing the likelihood function while moving trial point sources around in the region of the sky being analyzed.  
This approach has been driven by the limited photon counts and the relatively limited resolution of gamma-ray telescopes.  
However, at the sensitivity of the LAT, even a relatively "quiet" part of the sky may have 10 or more point sources close enough together to need to be modeled simultaneously when maximizing the (computationally expensive) likelihood function.  For this reason and because of the need to search in time, non-parametric algorithms for detecting sources are being investigated.

\subsection*{Literature overview for Poisson denoising using wavelets}
A host of estimation methods have been proposed in the literature \JF{for non-parametric Poisson noise removal}. Major contributions consist of \textbf{variance stabilization}: a classical solution is to preprocess the data by applying a variance stabilizing transform (VST) such as the Anscombe transform~\citep{rest:anscombe48}\citep{Donoho1993}. It can be shown that the transformed data are approximately stationary, \JF{independent}, and Gaussian. However, these transformations are only valid for a sufficiently large number of counts \SD{per pixel} (and of course, for even more counts, the Poisson distribution becomes Gaussian with equal mean and variance) \citep{starck:mur95_2}. 
The necessary average number of counts is about 20 if bias is to be avoided. 

In this case, as \SD{an} alternative approach, a filtering approach for very small numbers of 
counts, including frequent zero cases, has been proposed  in \citep{starck:sta98_1}, which is based on 
the popular isotropic \JF{undecimated} wavelet transform (\JF{implemented with the so-called} \`a trous algorithm) \citep{starck:book06} and the autoconvolution histogram technique for deriving the \JF{probability density function (pdf)} of the wavelet coefficient \citep{astro:slezak93,wave:jammal99,starck:book06}. 
This method is part of the data reduction pipeline of  the XMM-LSS  project \citep{pierre04} for   detecting of clusters of galaxies  \citep{pierre07}. This algorithm is obviously \JF{a good candidate} for {\textit{Fermi}} LAT 2D map analysis, but its extension to 2D-1D \JF{data sets} does not exist. It is far from being trivial, and even if it were possible, computation time would  certainly be prohibitive to \JF{allow} \SD{its use} for {\textit{Fermi}} LAT 2D-1D \JF{data sets}. Then, an alternative approach is needed. Several authors \citep{wave:kolac97,wave:timmermann99,wave:nowak99,wave:jammal99,rest:nason04,Zhang2007} 
have suggested that the Haar wavelet transform is
very well-suited for treating data with Poisson noise. 
Since a Haar wavelet coefficient
is just the difference between two random variables following a Poisson
distribution, it is easier to derive mathematical tools for removing  the 
noise than with any other wavelet method.  
\citet {starck:book06}  study shows  that the Haar transform is less effective for 
restoring X-ray astronomical images than the \`a trous algorithm. The reason is that the  wavelet shape
of the isotropic wavelet transform is much better adapted to astronomical sources, which are more or less Gaussian-shaped and isotropic, than the
Haar wavelet. Some papers \citep{scargle98,rest:kolaczyk04,rest:willet05,Willett06} proposed a  spatial partitioning, possibly dyadic,
of the image for complicated geometrical content recovery. This dyadic partitioning concept is however again  not very well suited to astrophysical data. 

\subsubsection*{\JF{The MSVST} alternative}
In a recent paper, \citet{starck:zhang07} have proposed to merge a variance stabilization technique and the multiscale decomposition, leading to the Multi-Scale Variance Stabilization Transform (\JF{MSVST}). In the case of the isotropic undecimated wavelet transform, as the wavelet coefficients $w_j$ \JF{are} derived by a simple difference of two consecutive dyadic scales of the input image (see section~\ref{subsec:iuwt}), $\JF{w_j = a_{j-1} - a_{j}}$, the stabilized wavelet coefficients are obtained by applying a stabilization on both $\JF{a_{j-1}}$ and  $\JF{a_{j}}$,  $\JF{w_j=  {\cal A}_{j-1} (a_{j-1})  - {\cal A}_{j} ( a_{j})}$, where $ {\cal A}_{j-1} $ and $ \JF{{\cal A}_{j}} $ are non-linear transforms \JF{that can be seen as a generalization of} the Anscombe transform; \JF{see section~\ref{sect_msvst} for details}. This new method is fast and easy to implement, and more \JF{importantly}, works very well at very low count \JF{situations}, down to $0.1$ photons per \JF{pixel}.

\subsubsection*{This paper}
In this paper,  we present a new multiscale \JF{representation}, derived from the \JF{MSVST}, which allows us to remove the Poisson noise in 3D data \JF{sets}, when the third dimension is not a spatial dimension, but the wavelength, the energy \JF{or} the time. \JF{Such 3D data are called 2D-1D data sets in the sequel}. We show that it could be very useful to analyze {\textit{Fermi}} LAT data, especially \JF{when looking for rapidly} time varying sources. Section~\ref{sec_glast_data}  describes the {\textit{Fermi}} LAT simulated data. Section~\ref{sect_msvst} reviews the \JF{MSVST} method relative to the isotropic undecimated wavelet transform and section~\ref{sect_msvst2d1d} shows how it can be extended to the 2D-1D case. Section~\ref{sect_glast_exp} presents some experiments on \SD{simulated {\textit{Fermi}} LAT data}. Conclusions are given in section~\ref{sect_ccl}.

 
\subsection*{Definitions and notations}
\label{subsec:notations}
For a real discrete-time filter whose impulse response is $h[i]$, $\bar{h}[i]=h[-i], ~ i \in \mathbb{Z}$ is its time-reversed version.
For the sake of clarity, the notation $h[i]$ is used instead of $h_i$ for the location index. This will lighten the notation by avoiding multiple subscripts 
in the derivations of the paper. The discrete circular convolution product of two signals will be written $\star$, and the continuous convolution of two functions $*$. The term circular stands for periodic boundary conditions. The symbol $\delta[i]$ is the Kronecker delta. 

For the octave band wavelet representation, analysis (respectively, synthesis) filters are denoted $h$ and $g$ 
(respectively, $\tilde{h}$ and $\tilde{g}$). The scaling and wavelet functions {used} for the analysis (respectively, synthesis) are denoted $\phi$ (with $\phi(\frac{x}{2}) = \sum_k h[k] \phi(x-k), x \in {\mathbb{R}} ~ and ~ k \in {\mathbb{Z}}$) and $\psi$ (with $\psi(\frac{x}{2}) = \sum_k g[k] \phi(x-k), x \in {\mathbb{R}} ~ and ~ k \in {\mathbb{Z}}$) (respectively, $\tilde{\phi}$ and $\tilde{\psi}$). We also define the scaled dilated and translated version of $\phi$ at scale $j$ and position $k$ as $\phi_{j,k}(x) = 2^{-j} \phi(2^{-j} x -k)$, and similarly for $\psi$, $\tilde{\phi}$ and $\tilde{\psi}$. A function $f(x,y)$ is isotropic if it is constant along all points $(x,y)$ that are equidistant from the origin.

A distribution is stabilized if its variance is made constant, typically equal to 1, independently of its mean. A transformation applied to a random variable is called a variance stabilizing transform (VST), if the distribution of the transformed variable is stabilized and is approximately Gaussian. 

\subsection*{Glossary}
{
\begin{tabular}{ll}
WT      &       Wavelet Transform \\
DWT     &       Discrete (decimated) Wavelet Transform \\
UWT     &       Undecimated Wavelet Transform \\
IUWT    &       Isotropic Undecimated Wavelet Transform \\
VST &  Variance Stabilization Transform \\
MSVST & Multi-Scale Variance Stabilization Transform \\
LAT    & Large Area Telescope (LAT) \\
FDR & False Discovery Rate
\end{tabular}}

\section{Data description}
\label{sec_glast_data}
\begin{figure}
\caption{Cutaway view of the LAT.  The LAT is modular; one of the 16 towers is shown with its tracking planes revealed.  High-energy gamma rays convert to electron-positron pairs on tungsten foils in the tracking layers.  The trajectories of the pair are measured very precisely using silicon strip detectors in the tracking layers and the energies are determined with the CsI calorimeter at the bottom.  The array of plastic scintillators that cover the towers provides an anticoincidence signal for cosmic rays.  The outermost layers are a thermal blanket and micrometeoroid shield.  The overall dimensions are $1.8 \times 1.8 \times 0.75$ m. \label{fig:fig1}}
\end{figure}

\subsection{{\textit{Fermi}} Large area telescope}
The LAT (Fig.~\ref{fig:fig1}) is a photon-counting detector, converting gamma rays into positron-electron pairs for detection.  The trajectories of the pair are tracked and their energies measured in order to reconstruct the direction and energy of the gamma ray.

The energy range of the LAT is very broad, approximately 20 MeV -- 300 GeV.  At energies below a few hundred MeV, the reconstruction and tracking efficiencies are lower, and the angular resolution is poorer, than at higher energies. The point spread function (PSF) width varies from about 3.5$^\circ$ at 100 MeV to better than 0.1$^\circ$ (68\% containment) at 10 GeV and above.  Owing to large-angle multiple scattering in the tracker, the PSF has broad tails; the 95\%/68\% containment ratio may be as large as 3.  

Wavelet denoising of LAT data has application as part of an algorithm for quickly detecting celestial sources of gamma rays.  The fundamental inputs to high-level analysis of LAT data will be energies, directions, and times of the detected gamma rays.  (Pointing history and instrument live times are also inputs for exposure calculations.)  For the analysis presented here, we consider the LAT data for some range of time to have been binned into 'cubes' $v(x,y,t)$ of spatial coordinates and time or, $v(x,y,E)$ of spatial coordinates and energy, because, as we shall see, the wavelet denoising can be applied in multiple dimensions, and so permits estimation of counts spectra. The motivations for filtering data with Poisson noise in the wavelet domain are well known{\textemdash}sources of small angular size are localized in wavelet space.

\subsection{Simulated LAT data}
\label{subsec:latsimudata}
The application of \JF{MSVST} to problems of detection and characterization of LAT sources was investigated using simulated data. The simulations included a realistic observing strategy (sky survey with the proper orbital and rocking periods) and response functions for the LAT (effective area and angular resolution as functions of energy and angle). Point sources of gamma rays were defined with systematically varying fluxes, spectral slopes, and/or flare intensities and durations.  The simulations also included a representative level of diffuse 'background' (celestial plus residual charged-particle) for regions of the sky well removed from the Galactic equator, where the celestial diffuse emission is particularly intense. The denoising results reported in Section~\ref{sect_glast_exp} use a data cube obtained according to this simulation scenario.




\section{The 2D multiscale variance stabilization transform (\JF{MSVST}) }
\label{sect_msvst}
In this section, we review the \JF{MSVST} method \citep{starck:zhang07}, restricted to the Isotropic Undecimated Wavelet Transform (IUWT).
Indeed, the \JF{MSVST} can use other transforms such \JF{as} the standard \JF{three-orientation} undecimated wavelet transform, the ridgelet  or the curvelet transforms; see \citep{starck:zhang07}. In our specific case \JF{here}, only the \JF{IUWT} is of interest.

\subsection{VST of a filtered Poisson process}
Given $\fX$ a sequence of $n$ independent Poisson random variables $X_i, i=1,\cdots,n,$ each of mean $\lambda_i$, \JF{let} $Y_i = \sum_{j=1}^n h[j]X_{i-j}$ \JF{be} the filtered process obtained by convolving the sequence $\fX$ with a discrete filter $h$. $Y$ denotes any one of the $Y_i$'s, and $\tau_k = \sum_i (h[i])^k$ for $k=1,2,\cdots$. 

If $h=\delta$, then \JF{we recover} the Anscombe \JF{VST} \citep{rest:anscombe48} of $Y_i$ (hence $X_i)$ \JF{which} acts as if the \JF{stabilized} data arose from a Gaussian white noise with unit variance, under the assumption that the intensity $\lambda_i$ is large. This is why the Anscombe VST performs poorly in low-count settings. But, if the filter $h$ acts as an ``averaging'' kernel (more generally a low-pass filter), one can reasonably expect that stabilizing $Y_i$ would be more beneficial, since the signal-to-noise ratio measured at the output of $h$ is expected to be higher.

Using a local homogeneity assumption, i.e. $\lambda_{i-j}=\lambda$ for all $j$ within the support of $h$, it has been shown \citep{starck:zhang07}
that for a non-negative filter $h$, the transform $Z =  b \sqrt{Y + c}$  with $b > 0$ and $c >0$ defined as
\be
\label{eq:c}
c = \frac{7\tau_2}{8\tau_1} - \frac{\tau_3}{2\tau_2} \quad,\quad b =   2\sqrt{\frac{\tau_1}{\tau_2}}
\ee
is a second order accurate variance stabilization transform, with asymptotic unit variance. By second-order accurate, we mean that the error term in the variance of the stabilized variable $Z$ decreases rapidly as $O(\lambda^{-2})$. From \eqref{eq:c}, it is obvious that when $h=\delta$, we obtain the classical Anscombe VST parameters $b=2$ and $c=3/8$. The authors in \citep{starck:zhang07} have also proved that $Z$ is asymptotically distributed as a Gaussian variate with mean  $b \sqrt{\tau_1\lambda}$ and unit variance. A non-positive $h$ with a negative $c$ could also be considered; see \citep{starck:zhang07} for more details.

 \begin{figure*}[htb]
\caption{Behavior of the expectation \JF{$\bE[Z]$ (left) and variance $\var{Z}$ (right)} as a function of the underlying intensity, for the Anscombe VST, 2D Haar-Fisz VST, and out \JF{VST} with the 2D $B_3$-Spline filter as a low-pass filter $h$.}
\label{fig_msvst}
\end{figure*}

Fig.\ref{fig_msvst} shows the \JF{Monte-Carlo} estimates of \JF{the expectation $\bE[Z]$ (left) and the variance $\var{Z}$ (right)} obtained from $2 \cdot 10^5$ Poisson noise realizations of $\fX$, plotted as a function of the intensity $\lambda$ for both Anscombe \citep{rest:anscombe48} (dashed-dotted), Haar-Fisz (dashed)\citep{rest:nason04} and our VST with the 2D $B_3$-Spline filter as a low-pass filter $h$ (solid). The asymptotic bounds (dots) (i.e.~$1$ for the variance and $\sqrt{\lambda}$ for the expectation) are also shown. 
It can be seen that for increasing intensity, $\bE[Z]$ and $\var{Z}$ approach the theoretical bounds at different rates depending on the VST used.   
Quantitatively, Poisson variables transformed using the Anscombe VST can be reasonably considered to be unbiased and stabilized for $\lambda \gtrapprox 10$, using Haar-Fisz for $\lambda \gtrapprox 1$, and using out \JF{VST} (after low-pass filtering with the chosen $h$) for $\lambda \gtrapprox 0.1$. 


\subsection{The isotropic undecimated wavelet transform}
\label{subsec:iuwt}
The undecimated wavelet transform (UWT) uses an analysis filter bank $(h,g)$ to decompose a signal $a_0$ into a coefficient set $W = \{d_1, \dots, d_J, a_J\}$, where $d_j$ is the wavelet (detail) coefficients at scale $j$ and $a_J$ is the approximation coefficients at the coarsest resolution $J$. The passage from one resolution to the next one is obtained using the  
``\`a trous'' algorithm~\citep{Holschneider1989}\citep{Shensa1992}
\begin{eqnarray}
\label{eq:uwtdecomp}
a_{j+1}[l] &=& (\bar{h}^{\uparrow j} \star a_{j})[l] = \sum_k h[k] a_{j}[l+2^{j}k], \quad \\
w_{j+1}[l] &=& (\bar{g}^{\uparrow j} \star a_{j})[l] = \sum_k g[k] a_{j}[l+2^{j}k] ,
\end{eqnarray}
where $h^{\uparrow j}[l] = h[l]$ if $l / 2^j \in \bZ$ and $0$ otherwise, $\bar{h}[l] = h[-l]$, and ``$\star$'' denotes discrete circular convolution. The reconstruction is given by $a_{j}[l] = \frac{1}{2}\left[ (\tilde{h}^{\uparrow j} \star a_{j+1})[l] + (\tilde{g}^{\uparrow j} \star w_{j+1})[l] \right]$. The filter bank $(h,g,\tilde{h},\tilde{g})$ needs to satisfy the so-called exact reconstruction condition \JF{\citep{ima:mallat98,starck:book06}}.

The Isotropic UWT (IUWT) \citep{Starck2006} uses the filter bank $(h,g=\delta-h,\tilde{h}=\delta,\tilde{g}=\delta)$
where $h$ is typically a symmetric low-pass filter such as the $B_3$-Spline filter. 
The reconstruction is trivial, i.e., $a_0 = a_J + \sum_{j=1}^J w_{j}$. This algorithm is widely used in astronomical applications~\citep{starck:book98} and biomedical imaging~\citep{OlivoMarin2002} to detect isotropic objects. 

The IUWT filter bank in $q$-dimension ($q\geq2$) becomes $(h_{q\text{D}},g_{q\text{D}}=\delta-h_{q\text{D}},\tilde{h}_{q\text{D}}=\delta,\tilde{g}_{q\text{D}}=\delta)$ where $h_{q\text{D}}$ is the tensor product of $q$ 1D filters $h_{\mathrm{1D}}$. Note that $g_{q\text{D}}$ is in general non-separable.

\subsection{\JF{MSVST} with the IUWT}
\label{subsec:msvst+iuwt}
Now the VST can be combined with the IUWT in the following way: since the filters $\bar{h}^{\uparrow j}$ at all scales $j$ are low-pass filters (so have nonzero means), we can first stabilize the approximation coefficients $a_j$ at each scale using the VST, and then compute in the standard way the detail coefficients from the stabilized $a_j$'s. Given the particular structure of the IUWT analysis filters $(h,g)$, the stabilization procedure is given by
\be
\label{eq:coupled:msvst}
&\mbox{IUWT}&
\left\{\begin{array}{lll} 
a_j &=& \bar{h}^{\uparrow j-1} \star a_{j-1} \\
w_j &=& a_{j-1} - a_j
\end{array}\right. \nonumber\\ \Longrightarrow  
&\begin{tabular}{c}
\mbox{\JF{MSVST}} \\
\mbox{+} \\
\mbox{IUWT}
\end{tabular} &
\left\{\begin{array}{lll}
a_j &=& \bar{h}^{\uparrow j-1} \star a_{j-1} \\
w_j &=& {\cal{A}}_{j-1}(a_{j-1}) -  {\cal{A}}_j(a_j)
\end{array}\right. ~.
\ee
Note that the VST is now scale-dependent (hence the name \JF{MSVST}). The filtering step on $a_{j-1}$ can be rewritten as a filtering on $a_0=\fX$, i.e., $a_{j} = h^{(j)}\star a_0$, where $h^{(j)} = \bar{h}^{j-1}\star\cdots\star\bar{h}^{1}\star \bar{h}$ for $j\geq 1$ and $h^{(0)} = \delta$. ${\cal{A}}_j$ is the VST operator at scale $j$
\begin{equation}
\label{eq:vstj:separated}
{\cal{A}}_j(a_j) = b^{(j)}  \sqrt{a_j + c^{(j)}} ~.
\end{equation}

Let us define $\tau_k^{(j)} = \sum_i \parenth{h^{(j)}[i]}^k$.
Then according to (\ref{eq:c}), the constants $b^{(j)}$ and $c^{(j)}$ associated to $h^{(j)}$ must be set to 
\begin{equation}
\label{eq:cj:separated}
c^{(j)} = \frac{7\tau_2^{(j)}}   {8\tau_1^{(j)}} - \frac{\tau_3^{(j)}} {2\tau_2^{(j)}} \quad,\quad b^{(j)} =   2\sqrt{\frac{\tau_1^{(j)}}{\tau_2^{(j)}}} ~.
\end{equation}
The constants $b^{(j)}$ and $c^{(j)}$ only depend on the filter $h$ and the scale level $j$. They can all be pre-computed once for any given $h$. 
A schematic overview of the decomposition and the inversion of \JF{MSVST}+IUWT is depicted in Fig.~\ref{fig:msvst}.

In summary, IUWT denoising with the \JF{MSVST} involves the following three main steps:
\begin{enumerate}
\item {\bf Transformation} : Compute the \JF{IUWT} in conjunction with the \JF{MSVST} as described above.

\item {\bf Detection} : Detect significant detail coefficients by hypothesis testing. The appeal of a binary hypothesis testing approach is that it allows quantitative control of significance. Here, we take benefit from the asymptotic Gaussianity of the stabilized $a_j$'s that will be transferred to the $w_j$'s as it has been shown by \citep{starck:zhang07}. Indeed, these authors have proved that under the null hypothesis $H_0: w_j[k] = 0$ corresponding to the fact that the signal is homogeneous (smooth), the stabilized detail coefficients $w_j$ follow asymptotically a centered normal distribution with an intensity-independent variance; \JF{see \citep[Theorem 1]{starck:zhang07} for details}. This variance depends only on the filter $h$ and the current scale, and can be tabulated once for any $h$. Thus, the distribution of the $w_j$'s being known (Gaussian), we can detect the significant coefficients by classical binary hypothesis testing.

\item {\bf Estimation} : Reconstruct the final estimate using the knowledge of the detected coefficients. This step requires inverting the \JF{MSVST} after the detection step. For the IUWT filter bank, there is a closed-form inversion expression as we have 
\begin{equation}
\label{eq:iuwt:inverse}
\JF{
a_0 = {\cal{A}}_0^{-1}\left[{\cal{A}}_J(a_J) + \sum_{j=1}^J w_j\right] ~.
}
\end{equation} 
\end{enumerate}

\begin{figure*}[htb]
\caption{Diagrams of the \JF{MSVST} combined with the IUWT. The notations are the same as those of \eqref{eq:coupled:msvst} and \eqref{eq:iuwt:inverse}. The left dashed frame shows the decomposition part. Each stage of this frame corresponds to a scale $j$ and an application of \eqref{eq:coupled:msvst}. The right dashed frame illustrates the direct inversion \eqref{eq:iuwt:inverse}.}
\label{fig:msvst}
\end{figure*} 


\subsubsection{Example}


\begin{figure*}[htb]
\centerline{
\hbox{
\hbox{
}
}
}
\caption{Top, XMM simulated data, and Haar-Kolaczyk \JF{\citep{wave:kolac97}} filtered image.
Bottom, Haar-Jammal-Bijaoui \JF{\citep{wave:jammal99}} and \JF{MSVST} filtered images. Intensities 
logarithmically transformed.}
\label{fig_xmm_simu_roset}
\end{figure*}


Fig.~\ref{fig_xmm_simu_roset} upper left shows a set of objects of different sizes and different intensities \JF{contaminated by} a Poisson noise. Each object along any radial branch has the same integrated intensity \JF{within its support} and has a more and more extended support as we go farther from the center. The integrated intensity reduces as the branches turn in the clockwise direction. Denoising such an image is challenging. Fig.~\ref{fig_xmm_simu_roset}, top-right, bottom-left and right, show 
respectively the filtered images by Haar-Kolaczyk \JF{\citep{wave:kolac97}}, Haar-Jammal-Bijaoui \JF{\citep{wave:jammal99}} and
the \JF{MSVST}.  



As expected, the relative merits (sensitivity) of the \JF{MSVST} estimator become increasingly salient as we go farther from the center, and as the branches turn clockwise. That is, the \JF{MSVST} estimator outperforms its competitors as the intensity becomes low. Most sources were detected by the \JF{MSVST} estimator even for very low counts situations; see the last branches clockwise in Fig.~\ref{fig_xmm_simu_roset} bottom right and compare to Fig.~\ref{fig_xmm_simu_roset} top right and Fig.~\ref{fig_xmm_simu_roset} bottom left.

\section{\JF{2D-1D} \JF{MSVST} denoising}
\label{sect_msvst2d1d}

\subsection{2D-1D wavelet transform}
In the previous section, we have seen how a Poisson noise can be removed from 2D image using the \JF{IUWT} 
and the \JF{MSVST}. Extension to a \JF{$q$D} data \JF{sets} is straightforward, and the 
denoising will be nearly optimal as long as \JF{each object} belonging to this \JF{$q$-dimensional} space \JF{is} roughly isotropic. 
In the case of 3D data where the third dimension is either the time or the energy, we are clearly not in this configuration, \JF{and the naive analysis of a 3D isotropic wavelet does not make sense}.
Therefore, we want to analyze the data with a \JF{non-isotropic} wavelet, where the \JF{time or energy} scale is not connected 
to the spatial scale. Hence, an ideal wavelet function would be defined by:
\begin{eqnarray}
\psi(x,y,z) = \psi^{(xy)}(x,y) \psi^{(z)}(z) ~,
\end{eqnarray}
where $\psi^{(xy)}$ is the spatial wavelet and $\psi^{(z)}$ \JF{is} the temporal \JF{(or energy)} wavelet.
In the following, we will consider only \JF{isotropic and dyadic spatial scales}, and we 
note $j_1$ the spatial \JF{resolution} index (i.e. scale = $2^{j_1}$), $j_2$ the time \JF{(or energy)} resolution index. \JF{Thus, define the scaled spatial and temporal (or energy) wavelets} 
\begin{align*}
&\psi_{j_1}^{(xy)}(x,y) = \frac{1}{2^{j_1}} \psi^{(xy)}   ( \frac{x}{2^{j_1}}, \frac{y}{2^{j_1}}) \text{ and } \\
&\psi_{j_2}^{(z)}(z) = \frac{1}{ \sqrt{2^{j_2}}} \psi^{(z)}   (\frac{z}{2^{j_2}}).
\end{align*}

Hence, we derive the wavelet coefficients
$w_{j_1,j_2} [k_x, k_y, k_z]$ from a given data set $D$ 
($k_x$ and $k_y$ are spatial index and $k_z$ a time (\JF{or} energy) index). In continuous coordinates, this amounts to the formula
\begin{eqnarray}
\label{eqn_wave_2d1d}
w_{j_1, j_2}[k_x, k_y, k_z]  & = & \frac{1}{2^{j_1}}   \frac{1}{\sqrt{2^{j_2}}}  \iiint_{-\infty}^{+\infty} D(x,y,z)    \nonumber \\
 & &  \hspace{2cm} {\psi^{(xy)} }  \left( \frac{x-k_x}{2^{j_1}}, \frac{y-k_y}{2^{j_1}}    \right) 
      {\psi^{(z)} } \left(  \frac{z-k_z}{2^{j_2}}  \right) dx dy dz  \nonumber  \\ 
  & = &  D * {\bar{\psi}}^{(xy)}_{j_1} * \bar{\psi}^{(z)}_{j_2} (x,y,z) ~,
\end{eqnarray}
where $*$ is the convolution and $\bar{\psi}(x) = \psi(-x)$.

\subsubsection*{Fast undecimated 2D-1D decomposition/reconstruction}
In order to have a fast algorithm for \JF{discrete data}, we \JF{use wavelet functions associated to filter banks}.
Hence, our wavelet decomposition consists in applying first a 2D \JF{IUWT} for each frame $k_z$.
Using the 2D \JF{IUWT}, we have \JF{the reconstruction formula}:
\begin{eqnarray}
\JF{
D[k_x,k_y,k_z] = a_{J_1}[k_x,k_y] + \sum_{j_1=1}^{J_1} w_{j_1}[k_x,k_y,k_z], ~ \forall k_z ~ ,
}
\end{eqnarray}
where $J_1$ is the number of spatial scales.
Then, for each \JF{spatial location} $\JF{(k_x,k_y)}$ and for each 2D wavelet scale scale $j_1$, we \JF{apply} a 1D wavelet transform along $z$ on the spatial wavelet coefficients $w_{j_1}[k_x,k_y,k_z]$ \JF{such that}
\begin{eqnarray}
w_{j_1}[k_x,k_y,k_z] = w_{j_1, J_2} [k_x,k_y,k_z] + \sum_{j_2 = 1}^{J_2} w_{j_1,j_2} [k_x,k_y,k_z], ~ \forall (k_x,k_y) ~,
\end{eqnarray}
where $J_2$ is the number of scales along $z$.
The same processing is also applied on the coarse spatial scale $\JF{a_{J_1}[k_x,k_y,k_z]}$, and we have
\begin{eqnarray}
\JF{
a_{J_1}[k_x,k_y,k_z] = a_{J_1, J_2}[k_x,k_y,k_z]  + \sum_{j_2=1}^{J_2} w_{J_1,j_2} [k_x,k_y,k_z], ~ \forall (k_x,k_y) ~.
}
\end{eqnarray}

Hence, we have a 2D-1D undecimated wavelet representation of the input data $D$:
\begin{align}
&D[k_x,k_y,k_z] = a_{J_1,J_2}[k_x,k_y,k_z] +\sum_{j_1=1}^{J_1} w_{j_1,J_2} [k_x,k_y,k_z] + \nonumber\\
&\qquad\quad\sum_{j_2=1}^{J_2} w_{J_1,j_2} [k_x,k_y,k_z] + \sum_{j_1=1}^{J_1}  \sum_{j_2=1}^{J_2} w_{j_1,j_2} [k_x,k_y,k_z] ~.
\end{align}

\JF{From this expression,} we distinguish four kinds of coefficients:
\begin{itemize}
\item Detail-Detail \JF{coefficients ($j_1 \leq J_1$ and $j_2 \leq J_2$)}: 
\begin{align}
&w_{j_1,j_2}[k_x,k_y,k_z]  =  (\delta - \bar{h}_{\mathrm{1D}}) \star  \nonumber\\
&\parenth{h^{(j_2-1)}_{\mathrm{1D}} \star   a_{j_1-1}[k_x,k_y,.]   - h^{(j_2-1)}_{\mathrm{1D}}  \star a_{j_1}[k_x,k_y,.]} ~.
\end{align}
\item Approximation-Detail \JF{coefficients ($j_1 = J_1$ and $j_2 \leq J_2$)}: 
\begin{eqnarray}
w_{J_1,j_2}[k_x,k_y,k_z] =  h^{(j_2-1)}_{\mathrm{1D}} \star a_{J_1}[k_x,k_y,.] -  h^{(j_2)}_{\mathrm{1D}} \star a_{J_1}[k_x,k_y,.] ~.
\end{eqnarray}
\item Detail-Approximation \JF{coefficients ($j_1 \leq J_1$ and $j_2 = J_2$)}: 
\begin{eqnarray}
w_{j_1,J_2}[k_x,k_y,k_z] =    h^{(J_2)}_{\mathrm{1D}} \star a_{j_1-1}[k_x,k_y,.]  - h^{(J_2)}_{\mathrm{1D}} \star \JF{a_{j_1}}[k_x,k_y,.] ~.
\end{eqnarray}
\item Approximation-Approximation \JF{coefficients ($j_1 = J_1$ and $j_2 = J_2$)}: 
\begin{eqnarray}
\JF{
a_{J_1,J_2}[k_x,k_y,k_z] =   h^{(J_2)}_{\mathrm{1D}} \star a_{J_1}[k_x,k_y,.] ~.
}
\end{eqnarray}
\end{itemize}

As the 2D-1D \JF{undecimated wavelet transform just described} is fully linear, a Gaussian noise remains Gaussian after transformation. Therefore, all thresholding strategies which have been developed for wavelet Gaussian denoising are still valid with the 2D-1D wavelet transform. Denoting $\mathrm{TH}$ the thresholding operator, the denoised cube in the case of additive white Gaussian noise is obtained by:
\begin{align}
&{\tilde D}[k_x,k_y,k_z] = a_{J_1,J_2}[k_x,k_y,k_z]  + \sum_{j_1=1}^{J_1} \mathrm{TH}( w_{j_1,J_2} [k_x,k_y,k_z]) \nonumber\\
&+ \sum_{j_2=1}^{J_2}  \mathrm{TH}( w_{J_1,j_2} [k_x,k_y,k_z] ) + \sum_{j_1=1}^{J_1}  \sum_{j_2=1}^{J_2}  \mathrm{TH}(  w_{j_1,j_2} [k_x,k_y,k_z] ) ~.
\end{align}
A typical choice of  $\mathrm{TH}$ is the hard thresholding operator, i.e. $\JF{\mathrm{TH}}(x) = 0$ if $ | x | $ \JF{is} below a given threshold $\tau$, and $\mathrm{TH}(x) = x$ \JF{if} $ | x | \ge \tau $. The threshold $\tau$ is generally \JF{chosen} between 3 and 5 times the noise standard deviation \citep{starck:book06}.

\subsection{Variance stabilization}
\label{subsec:2D1Dmsvst}

Putting all pieces together, we are now ready to plug the \JF{MSVST} into the 2D-1D \JF{undecimated} wavelet transform. Again, we distinguish four \JF{kinds} of coefficients that take the following \JF{forms}:
\begin{itemize}
\item Detail-Detail \JF{coefficients ($j_1 \leq J_1$ and $j_2 \leq J_2$)}: 
\begin{align}
\label{eq:detdet}
&w_{j_1,j_2}[k_x,k_y,k_z]  = (\delta - \bar{h}_{\mathrm{1D}}) \star \bigg(  \mathcal{A}_{j_1-1,j_2-1}\bigg[ h^{(j_2-1)}_{\mathrm{1D}} \star  \nonumber \\
&\quad a_{j_1-1}[k_x,k_y,.]\bigg] - \mathcal{A}_{j_1,j_2-1}    \crochets{ h^{(j_2-1)}_{\mathrm{1D}}    \star a_{j_1}   [k_x,k_y,.]  }  \bigg)  ~.
\end{align}
The schematic overview of the way the detail coefficients \JF{$w_{j_1,j_2}$} are computed is illustrated in \JF{Fig.}~\ref{fig:multiband}.
\item Approximation-Detail \JF{coefficients ($j_1 = J_1$ and $j_2 \leq J_2$)}: 
\begin{align}
\label{eq:approxdet}
w_{J_1,j_2}[k_x,k_y,k_z] & =  \mathcal{A}_{J_1,j_2-1}\crochets{h^{(j_2-1)}_{\mathrm{1D}} \star a_{J_1}[k_x,k_y,.]} - \nonumber \\ &\mathcal{A}_{J_1,j_2}\crochets{h^{(j_2)}_{\mathrm{1D}} \star a_{J_1}[k_x,k_y,.]} ~.
\end{align}
\item Detail-Approximation \JF{coefficients ($j_1 \leq J_1$ and $j_2 = J_2$)}: 
\begin{align}
\label{eq:detapprox}
w_{j_1,J_2}[k_x,k_y,k_z] & = \mathcal{A}_{j_1-1,J_2}\crochets{h^{(J_2)}_{\mathrm{1D}} \star a_{j_1-1}[k_x,k_y,.]} - \nonumber \\ &\JF{\mathcal{A}_{j_1,J_2}}\crochets{h^{(J_2)}_{\mathrm{1D}} \star \JF{a_{j_1}}[k_x,k_y,.]} ~.
\end{align}
\item Approximation-Approximation \JF{coefficients ($j_1 = J_1$ and $j_2 = J_2$)}: 
\begin{eqnarray}
\label{eq:approxapprox}
  c_{J_1,J_2}[k_x,k_y,k_z] =   h^{(J_2)}_{\mathrm{1D}} \star a_{J_1}[k_x,k_y,.] ~.
\end{eqnarray}
\end{itemize}

Hence, all \JF{2D-1D wavelet coefficients $w_{j_1,j_2}$ are now stabilized, and the noise on all these wavelet coefficients is Gaussian with known scale-dependent variance that depends solely on $h$}. Denoising is however not straightforward because there is no explicit reconstruction \JF{formula available} because of the form of the stabilization equations above. Formally, the stabilizing operators $\mathcal{A}_{j_1,j_2}$ and the convolution operators along $(x,y)$ and $z$ do not commute, even though the filter bank satisfies the exact reconstruction formula. \JF{To circumvent this difficulty, we propose to solve this reconstruction problem} by defining the multiresolution support \citep{starck:mur95_2} from the stabilized coefficients, and by using an iterative reconstruction scheme. 

\begin{figure*}[htb]
\caption{Overview of \JF{MSVST} with the 2D-1D IUWT. The diagram summarizes the main steps for computing the detail coefficients $w_{j_1,j_2}$ in \eqref{eq:detdet}. The notations are exactly the same as those of subsection~\ref{subsec:2D1Dmsvst} with $\bar{g}_{\mathrm{1D}}=\delta-\bar{h}_{\mathrm{1D}}$.}
\label{fig:multiband}
\end{figure*}

\subsection{Detection-reconstruction}
As the noise on the stabilized coefficients is Gaussian, \JF{and without loss of generality, we let its standard deviation equal to 1}, 
we consider that a wavelet coefficient $w_{j_1,j_2}[k_x,k_y,k_z]$ is significant, i.e., not due to noise, if 
its absolute value is larger than \JF{a critical threshold $\tau$, where $\tau$ is typically between 3 and 5}.  

The multiresolution support will be obtained by detecting at each scale the significant coefficients. 
The multiresolution support for \JF{$j_1 \le J$ and $j_2 \le J_2$} is defined as
\JF{
\begin{eqnarray} 
M_{j_1, j_2} [k_x,k_y,k_z] = 
\begin{cases}
1 & \text{ if $w_{j_1, j_2} [k_x,k_y,k_z]$  is significant,} \\ 
0 & \text{otherwise.} 
\end{cases} 
\end{eqnarray}}

In words, the multiresolution support $M$ indicates at which scales (spatial and \JF{time/energy}) and which positions, we have significant signal. We \JF{denote} ${\cal W}$ the 2D-1D \JF{undecimated wavelet transform described above}, $\JF{{\cal R}}$ the inverse wavelet transform and $Y$ the input \JF{noisy} data cube.
 
We want our solution $X$ \JF{to preserve the significant structures in the original data by reproducing} exactly the same coefficients as the wavelet coefficients of the input data $Y$, but only at scales and positions where significant signal has been detected (i.e. $ M {\cal W} X = M {\cal W} Y$). At other scales and positions, we want the smoothest solution \JF{with the lowest budget in terms of wavelet coefficients. Furthermore, as Poisson intensity functions are positive by nature, a positivity constraint is imposed on the solution}. It is clear that there are many solutions satisfying the positivity and multiresolution support consistency requirements, e.g. $Y$ itself. Thus, our reconstruction problem based solely on these constraints is an ill-posed inverse problem that must be regularized. Typically, the solution in which we are interested must be sparse by involving the lowest budget of wavelet coefficients. Therefore \JF{our reconstruction is formulated as a constrained sparsity-promoting minimization problem that can be written as follows}
\begin{align}
\min_{X} ~ \parallel {\cal  W} X \parallel_1 \quad \text{ subject to} \quad  & \begin{cases}
										M {\cal W} X = M {\cal W} Y \\
										\text{ and } X \geq 0
										\end{cases} ~ ,
\end{align}
\JF{where $\parallel . \parallel_1$ is the $\ell_1$-norm playing the role of regularization and is well known to promote sparsity \citep{Donoho2004a}. This problem can be solved efficiently using the hybrid steepest descent algorithm \citep{wave:yamada01,starck:zhang07}, and requires about 10 iterations in practice. Transposed into our context, its main steps can be summarized as follows:
\vspace{0.2cm}
\hrule
\begin{algorithmic}[1]
\REQUIRE Input noisy data $Y$; a low-pass filter $h$; multiresolution support $M$ from the detection step; number of iterations $N_{\max}$.\\
\STATE Initialize $X^{(0)}=M{\mathcal{W}}Y=M w_Y$,
\FOR{$t=1$ to $N_{\max}$}
\STATE $\tilde{d} =  Mw_Y + (1-M){\mathcal{W}}X^{(t-1)}$,
\STATE $X^{(t)}= P_+\left({\mathcal{R}}~\mathrm{ST}_{\beta_{t}}[\tilde{d}]\right)$,
\STATE Update the step $\beta_{t} = (N_{\max}-t)/(N_{\max}-1)$.
\ENDFOR
\end{algorithmic}
\hrule
\vspace{0.2cm}
where $P_+$ is the projector onto the positive orthant, i.e. $P_+(x) = \max(x,0)$. $\mathrm{ST}_{\beta_{t}}$ is the soft-thresholding operator with threshold $\beta_t$, i.e. $\mathrm{ST}_{\beta_{t}}[x] = x - \beta_{t}\mathrm{sign}(x)$ if $|x| \geq \beta_{t}$, and $0$ otherwise.}

\subsection{Algorithm summary}

The final \JF{MSVST} 2D-1D wavelet denoising algorithm is the following:
\JF{
\vspace{0.2cm}
\hrule
\begin{algorithmic}[1]
\REQUIRE Input noisy data $Y$; a low-pass filter $h$; threshold level $\tau$,~\\
\STATE {\bf \it \underline{2D-1D-MSVST}:} Apply the \JF{2D-1D-MSVST} to the data using \eqref{eq:detdet}-\eqref{eq:approxapprox}.
\STATE {\bf \it \underline{Detection}:} Detect the significant wavelet coefficients that are above $\tau$, and 
compute the multiresolution support $M$.
\item {\bf \it \underline{Reconstruction}:} Reconstruct the denoised data using the algorithm above.
\end{algorithmic}
\hrule}

\section{Experimental results and discussion}
\label{sect_glast_exp}

\subsection{\JF{MSVST-2D-1D versus MSVST-2D}}

\begin{figure*}[htb]
\centerline{
\vbox{
\hbox{
}
}}
\caption{Image obtained by \JF{integrating} along the $\JF{z}$-axis of the simulated data cube.}
\label{fig_grid}
\end{figure*}
We have simulated a data cube according to the procedure described in subsection~\ref{subsec:latsimudata}. The cube contains several sources, with spatial positions on a grid. It contains seven columns and five rows of LAT sources (i.e. 35 sources) with different power-law spectra. The cube size is $\JF{161 \times 161 \times 31}$, with a total number of photons equal \JF{to} $25948$, i.e. an average of $0.032$ photons per pixel.
Fig.~\ref{fig_grid} shows the 2D image obtained after integrating the simulated data cube along the $\JF{z}$-axis. Fig.~\ref{fig_cmp_msvst2d1d_msvst2d} shows a comparison between \JF{2D-MSVST} denoising of this image, and the image obtained by first applying a \JF{2D-1D-MSVST} denoising to the input cube, and integrating afterward along the $\JF{z}$-axis.
Fig.~\ref{fig_cmp_msvst2d1d_msvst2d} upper left and right \JF{show denoising results for the 2D-MSVST with respectively threshold values $\tau=3$ and $\tau=5$}, and Fig.~\ref{fig_cmp_msvst2d1d_msvst2d} bottom left and right \JF{show} the results for the \JF{2D-1D-MSVST} using respectively \JF{$\tau=4$ and $\tau=6$ detection levels}. \JF{The reason for using a higher threshold level for the 2D-1D cube is to correct for multiple hypothesis testings, and to get the same control over global statistical error rates. Roughly speaking, the number of false detections increases with the number of coefficients being tested simultaneously. Therefore, one must correct for multiple comparisons using e.g. the conservative Bonferroni correction or the false discovery rate (FDR) procedure \cite{Benjamini95}. As the number of coefficients is much higher with the whole 2D-1D cube, the critical detection threshold $\tau$ of 2D-1D denoising must be higher to have a false detection rate comparable to the 2D denoising}. 
As we can clearly see from Fig.~\ref{fig_cmp_msvst2d1d_msvst2d}, the results are very close. This means that applying a \JF{2D-1D} denoising on the cube instead of a 2D denoising on the integrated image does not degrade the detection power  of the \JF{MSVST}. The main advantage of the \JF{2D-1D-MSVST} is the fact that we recover the spectral (or temporal) information for each spatial position. Fig.~\ref{fig_msvst2d1d_frame} shows two frames (frame 16 top left and frame 25 bottom left) of the input cube and the same frames after the \JF{2D-1D-MSVST} denoising top right and bottom right. Fig.~\ref{fig_grid_spectra} \JF{displays} the obtained \JF{spectra} at two different spatial positions $(112,47)$ and $\JF{(126, 79)}$ which correspond to the centers of two distinct sources. 

\begin{figure*}[htb]
\centerline{
\vbox{
\hbox{
 }
\hbox{
}
}
}
\caption{Top, 2D-MSVST filtering on the integrated image with respectively a $\JF{\tau=3}$ and a $\JF{\tau=5}$ detection level.
Bottom, integrated image after a \JF{2D-1D-MSVST} denoising of the  simulated data cube, with respectively a $\JF{\tau=4}$ and a $\JF{\tau=6}$ detection level.}
\label{fig_cmp_msvst2d1d_msvst2d}
\end{figure*}

\begin{figure*}[htb]
\centerline{
\vbox{
\hbox{
 }
\hbox{
}
}
}
\caption{Top, frame number 16 of the input cube and the same frame after the 2D-1D-MSVST filtering at  $6\sigma$ .
Bottom,  frame number 25 of the input cube and the same frame after the 2D-1D-MSVST filtering at  $6\sigma$ .}
\label{fig_msvst2d1d_frame}
\end{figure*}

\begin{figure*}[htb]
\centerline{
\vbox{
\hbox{
}
}
}
\caption{Pixel spectra at two different spatial locations after the 2D-1D-MSVST filtering.}
\label{fig_grid_spectra}
\end{figure*}

\subsection{Time-varying source detection}

\begin{figure*}[htb]
\centerline{
\vbox{
\hbox{
 }
 }
 }
\caption{Time-varying source. From left to right, simulated source, temporal flux, and co-added image along the time axis of noisy data cube.}
\label{fig_time_source}
\end{figure*}

We have simulated a time varying source in a cube of size $\JF{64 \times 64 \times 128}$. The source has a Gaussian shape both in space and time. It is centered in the middle of the cube at $(32,32,64)$; i.e. its brightest point is at this location. The standard deviation of the Gaussian is 1.8 in space (pixel unit), and 1.2 along time (frame unit). The total flux of the source (i.e. spatial and temporal integration) is 100. We have added a background level of $0.1$.  Finally, Poisson noise was generated. \JF{Fig.}~\ref{fig_time_source} shows respectively from left to right an image of the original source, the flux per time frame and the integration of all \JF{noisy} frames along the time axis. As it can be seen, the source is hardly detectable in Fig.~\ref{fig_time_source} right. By running the \JF{2D-MSVST} denoising method on the time-integrated image, we were not able to detect it. Then we applied the \JF{2D-1D-MSVST} denoising method on the noisy 3D data set. This time, we were able to restore the source \JF{with a threshold level $\tau=6$}. Fig.~\ref{fig_rec_temp_source} left \JF{depicts} one frame (frame 64) of the denoised cube, and Fig.~\ref{fig_rec_temp_source} right shows the flux of the recovered source per frame (dotted line). The solid and thick-solid lines show respectively the flux per time frame after background subtraction in the noisy data and the original noise-free data set.  
We can conclude from this experiment that the \JF{2D-1D-MSVST} is able to recover \JF{rapidly time-varying} sources in \JF{the spatio-temporal data set, whereas even a robust algorithm such as the 2D-MSVST} method will completely fail if we integrate along the time axis. This was expected since the co-addition of all frames mixes the few frames containing the source with those which contain only the noisy background. Co-adding followed by a 2D detection is clearly suboptimal, except if we repeat the denoising procedure with many temporal windows with varying size. We can also notice that the \JF{2D-1D-MSVST} \JF{is able to recover} very well \JF{the times at which the source flares}, \JF{although} the source is slightly spread out on the time axis and  the flux of the source is not very well estimated, and other methods such \JF{as} maximum likelihood should be preferred for a correct flux estimation, once the sources have been detected.

\begin{figure*}[htb]
\centerline{
\hbox{
\hbox{
}
}
}
\caption{Recovered time-varying source. Left, one frame of the denoised cube. Right, flux per time frame for the noisy data after background subtraction (solid line), for the original noise-free cube (thick-solid line) and for the recovered source (dashed line).}
\label{fig_rec_temp_source}
\end{figure*}


\subsection{Diffuse emission of the Galaxy}

\begin{figure*}[htb]
\centerline{
\hbox{
}}
\caption{Left, from top to bottom, simulated data of the  diffuse gamma-ray emission of the Milky Way in energy band  171-181 Mev, 
noisy simulated data and filtered data using the MSVST.  Right, same images for energy band 9.87-1.04 GeV.}
\label{fig_frame_30_100}
\end{figure*}

In this experiment, we have simulated a $720 \times 360 \times 128$  cube using the Galprop code \cite{strong07} that has a model of the diffuse gamma-ray 
emission of the Milky Way. The units of the pixels are photons $cm^-{2} s^{-1} sr^{-1} MeV^{-1}$.
The gridding in Galactic  longitude and latitude is $0.5$ degrees, and the 128 energy planes are logarithmically
spaced from 30 MeV to 50 GeV. A six months LAT data set was created by multiplying the simulated cube with the exposure (6 months), and 
by convolving each energy band  with the point spread function of the LAT instrument. The PSF strongly varies with the energy.
Finally we have created the noisy observations assuming a Poisson noise distribution. 

Fig.~\ref{fig_frame_30_100} left shows from top to bottom the original simulated data, the noisy data and the filtered data for the band at energy 171-181 Mev. The same figures for the band 9.87-1.04 GeV are shown in Fig.~\ref{fig_frame_30_100} right.

\section{Conclusion}
\label{sect_ccl}
The motivations for a reliable nonparametric source detection algorithm to apply to {\textit{Fermi}} LAT data are clear.
Especially for the relatively short time ranges over which we will want to study sources, the data will be squarely in the low counts regime with widely varying response functions and significant celestial foregrounds.
In this paper, we have shown that the \JF{MSVST}, associated with a 2D-1D wavelet transform, is a very efficient way to detect time-varying sources. The proposed algorithm is as powerful as the \JF{2D-MSVST} applied to co-added frames to detect a source if the latter \JF{is slowly varying or constant over time}. But when the source is \JF{rapidly varying}, we lose some detection power when we co-add frames \JF{having no source and those containing the sources}. \JF{Our approach gives us an alternative to frame-co-adding and outperforms the 2D algorithms on the co-added frames. Unlike 2D denoising, our method fully exploits the information in the 3D data set and allows to recover the source dynamics by detecting temporally varying sources}.   

\begin{acknowledgements}
We thank Jean-Marc Casandjian for providing us  the simulated data set  of the diffuse emission of the Galaxy and Jeff Scargle for his helpful comments and critics.
This work was partially  supported by the French National Agency for Research (ANR -08-EMER-009-01).
 \end{acknowledgements}

\bibliographystyle{aa} 
\bibliography{JLSBibTex}

\end{document}